\documentstyle[12pt]{article}
\def\hybrid{\topmargin 0pt      \oddsidemargin 0pt
	\headheight 0pt \headsep 0pt
	\textheight 9in         
	\textwidth 6.25in       
	\marginparwidth .875in
	\parskip 5pt plus 1pt   \jot = 1.5ex}

\catcode`\@=11
\def\marginnote#1{}
\newcount\hour
\newcount\minute
\newtoks\amorpm
\hour=\time\divide\hour by60
\minute=\time{\multiply\hour by60 \global\advance\minute by-\hour}
\edef\standardtime{{\ifnum\hour<12 \global\amorpm={am}%
	\else\global\amorpm={pm}\advance\hour by-12 \fi
	\ifnum\hour=0 \hour=12 \fi
	\number\hour:\ifnum\minute<10 0\fi\number\minute\the\amorpm}}
\edef\militarytime{\number\hour:\ifnum\minute<10 0\fi\number\minute}

\def\draftlabel#1{{\@bsphack\if@filesw {\let\thepage\relax
   \xdef\@gtempa{\write\@auxout{\string
      \newlabel{#1}{{\@currentlabel}{\thepage}}}}}\@gtempa
   \if@nobreak \ifvmode\nobreak\fi\fi\fi\@esphack}
	\gdef\@eqnlabel{#1}}
\def\@eqnlabel{}
\def\@vacuum{}
\def\draftmarginnote#1{\marginpar{\raggedright\scriptsize\tt#1}}

\def\draft{\oddsidemargin -.5truein
	\def\@oddfoot{\sl preliminary draft \hfil
	\rm\thepage\hfil\sl\today\quad\militarytime}
	\let\@evenfoot\@oddfoot \overfullrule 3pt
	\let\label=\draftlabel
	\let\marginnote=\draftmarginnote
   \def\@eqnnum{(\theequation)\rlap{\kern\marginparsep\tt\@eqnlabel}%
\global\let\@eqnlabel\@vacuum}  }

\def\numberbysection{\@addtoreset{equation}{section}
	\def\theequation{\thesection.\arabic{equation}}}

\def\underline#1{\relax\ifmmode\@@underline#1\else
	$\@@underline{\hbox{#1}}$\relax\fi}

\def\titlepage{\@restonecolfalse\if@twocolumn\@restonecoltrue\onecolumn
     \else \newpage \fi \thispagestyle{empty}\c@page\z@
	\def\thefootnote{\fnsymbol{footnote}} }

\def\endtitlepage{\if@restonecol\twocolumn \else  \fi
	\def\thefootnote{\arabic{footnote}}
	\setcounter{footnote}{0}}  
\catcode`@=12
\relax
\def\ie{\hbox{\it i.e.}}        \def\etc{\hbox{\it etc.}}
\def\eg{\hbox{\it e.g.}}        \def\cf{\hbox{\it cf.}}
\def\etal{\hbox{\it et al.}}
\def\tr{\mathop{\rm tr}}
\def\Tr{\mathop{\rm Tr}}
\def\beq{\begin{equation}}
\def\eeq{\end{equation}}
\def\bea{\begin{eqnarray}}
\def\eea{\end{eqnarray}}
\def\bar{\overline}
\def\z{\bar {z}}
\def\nn{\nonumber}
\def\pa{\partial}
\def\d{{\cal D}}
\def\I{{\cal I}}
\def\t{{\theta}}
\def\s{\sigma}

\def\pz{\partial_z}
\def\pb{\partial_{\z}}
\def\pt{\partial_\t}

\def\m{\mu_{\z}^z}
\def\a{\alpha_{\z}^\t}
\def\c{c^z}
\def\g{\gamma^{\t}}
\def\b{b_{zz}}
\def\B{\beta_{z\t}}

\def\p{\Psi_{\z}^z}
\def\f{\Phi^z}
\def\F{\bar\Phi-{z\z}}
\def\n{\bar\eta-{z\z}}

\def\e{\epsilon ^z}
\def\k{\kappa ^\t}
\def\demi{{1\over 2}}
\def\quart{{1\over 4}}
\def\C{C^z}
\def\M{M^z}
\def\MM{\M+\C}
\def\dd{d+s}
\def\BB{B_{z\t}}
\def\l{l^z}
\def\ll{\lambda^\t}
\def\d{\delta}
\def\trois{{3\over 2}}
\def\L{{\Lambda+{z}}}
\def\TT{{T_{zz}}}
\def\GG{G_{z\t}}
\def\KK{K_{z\t}}

\def\ba{{\bar\alpha}}

\def\ee{\eea}
\def\be{\bea}

\def\eq{\eeq}
\def\bq{\beq}

\relax
\numberbysection
\hybrid
\begin{document}
\begin{titlepage}
\begin{center}

{\large\bf
Transmutation of Pure 2-D Supergravity Into Topological
2-D Gravity and Other Conformal Theories}\\[1in]
        {\bf    Laurent Baulieu}\footnote{email: baulieu@lpthe.jussieu.fr
\hfill    PAR--LPTHE 92--22, hep-th/9206019}\\
    	  LPTHE, Universit\'e Pierre et Marie Curie\\
{	Tour 16,
	4 place Jussieu,
 75005 Paris, France}\\
\end{center}

\vskip 1cm

\begin{quotation}
{\bf Abstract }
We consider the BRST and superconformal properties of    the ghost
action of 2-D supergravity. Using the  background spin
structure on the  worldsheet, we show that this action
 can be transformed  by canonical field transformations   to reach
other   conformal models such as the 2-D
topological gravity  or the chiral models for which the  gauge variation of the
action reproduces the left or right conformal anomaly.  Our method consists in
using the gravitino and its ghost
 as fundamental blocks to build fields with different conformal weights and
statistics. This indicates in particular that the twisting of a
conformal model into another one can be classically
 interpreted
as a change of "field representation" of the superconformal symmetry.

\end{quotation}
 \end{titlepage}

\def\ie{\hbox{\it i.e.}}
\def\etc{\hbox{\it etc.}} \def\eg{\hbox{\it
e.g.}}        \def\cf{\hbox{\it cf.}} \def\etal{\hbox{\it et al.}}
\def\tr{\mathop{\rm tr}} \def\Tr{\mathop{\rm Tr}} \def\beq{\begin{equation}}
\def\eeq{\end{equation}} \def\bea{\begin{eqnarray}} \def\eea{\end{eqnarray}}
\def\bar{\overline} \def\z{{\bar {z}}} \def\nn{\nonumber} \def\pa{\partial}
\def\d{{\cal D}} \def\I{{\cal I}} \def\t{{\theta}} \def\s{\sigma}

\def\pz{\partial_z} \def\pb{\partial_{\z}} \def\pt{\partial_\t}

\def\m{\mu_{\z}^z} \def\a{\alpha_{\z}^\t} \def\c{c^z} \def\g{\gamma^{\t}}
\def\b{b_{zz}} \def\B{\beta_{z\t}}

\def\p{\Psi_{\z}^z} \def\f{\Phi^z} \def\F{\bar\Phi-{z\z}}
\def\n{\bar\eta-{z\z}}

\def\e{\epsilon ^z} \def\k{\kappa ^\t} \def\demi{{1\over 2}} \def\quart{{1\over
4}} \def\C{C^z} \def\M{M^z} \def\MM{\M+\C} \def\dd{d+s} \def\BB{B_{z\t}}
\def\l{l^z} \def\ll{\lambda^\t} \def\d{\delta} \def\trois{{3\over 2}}
\def\L{{\Lambda^{z}}} \def\TT{{T_{zz}}} \def\GG{G_{z\t}} \def\KK{K_{z\t}}

\def\mo{{\m}_0} \def\ao{{\a}_0}

\def\ba{{\bar\alpha}}

\def\ee{\eea} \def\be{\bea}

\def\eq{\eeq} \def\bq{\beq}

\newpage

 \section{ Introduction}

Many conformal field theories are based
on  energy momentum tensors which, at the classical level, are not more than
quadratic in  the dynamical fields
 and their first and second order
derivatives. Their action is essentially a free one and can
be often  written as a superposition of $b-c$ systems. In
various cases a relationship   occurs  between the quantum energy momentum
tensor
operators of different theories. They can be "twisted" one into each
other by the addition of the derivative of an abelian current \cite{egu},
 and moreover
an $N=2$ supersymmetry is present \cite{fms}. Here, we
investigate whether these phenomena rely on  classical properties,
indepently on the quantization of  the fields. We consider
 conformal theories with equal number of commuting and anticommuting
fields, which is a necessary condition to get some
 kind of supersymmetry. Observing that the
only meaningful differences which can exist  at the classical level between
different models  are  the  conformal weights and statistics of the
fields, we find it  reasonable to believe that   a root theory should exist,
 where the fields   belong to what one could call the "fundamental
representation", the other "representations" being obtainable by suitable
canonical field transformations.  Since the conformal symmetry is supposed to
be
maintained in any "representation",  the     Lagrangians,  energy momentum
tensors and other conserved currents such as the ghost number and BRST Noether
currents should be also related  by these   changes of field variables. The
Ward identity expressing the conformal symmetry should have the same expression
in all representations, the only freedom in a given theory
being the value of the
coefficient  of the possible anomalous term. It is of course understandable
 that the anomaly coefficient
escapes the classical property that systems which are related by  canonical
field transformations are  equivalent, since its value must be computed at the
quantum level and involves a regularisation. One also expects that for two
different "representations" the  expressions of physical observables,  computed
at the level of field polynomials from the BRST cohomology of the theory or
from BRST exact terms,   be related by the canonical changes of variables, but
the interpretations and values of their expectation values, as the anomaly
coefficient,   differ when one goes from one theory to the other.

Our  point of view forces us to work with a non trivial reference metric on the
worldsheet, and moreover to introduce  a gravitino, in order to have a spin
structure on the worldsheet. Indeed, we find it natural  to use the gravitino
ghost field, which is a commuting object  with conformal weight one half, as a
building block to generate    by field multiplications  higher order conformal
field "representations", and  the  background   gravitino field, which is an
anticommuting   object, to possibly change the field statistics. The conformal
gauge results are easily obtained by setting the Weyl independent
parts of the background metric and  gravitino fields equal to zero. The use of
metric and  gravitino background fields has also the  advantage of simplifying
the   derivation of the various properties of the energy momentum and
supersymmetry tensors, since they they are the sources of these objects
\cite{bbgi} \cite{stora}.

The paper is organized as follows. We study  the ghost action of 2-D
supergravity in the gauges where the metric  and gravitino are conformally set
equal to non vanishing background metric and  gravitino fields, a particular
case
of which is the conformal gauge. In addition to the ordinary supergravity BRST
symmetry, present by construction, we find a background local supersymmetry,
acting on the supergravity ghost, antighost and background metric  and
gravitino
fields, with a generator which anticommutes with the basic BRST and ghost
number
symmetries.  In both holomorphic and antiholomorphic sectors, the formulae
 fall
in a very simple $N=1$ superfield formalism, although the background is non
trivial. The N=2 conformal supersymmetry \cite{fms} of the ghost action appears
as an accident of the case when the background gravitino is set equal to zero
and the Beltrami parameter is choosen constant. Indeed, with these choices of
the
background gauge, the
ghost current can be splitted into two independently conserved abelian
currents and this
  provides the  additional U(1) symmetry which is necessary to
extend the fundamental background N=1 supersymmetry into an N=2 conformal
supersymmetry. Then comes our main observation. The presence of the background
gravitino and of its ghost permits one to do canonical  field redefinitions,
 which
 change the conformal weights   as well as
the  statistics. The redefined
fields can still be seen as "realizing" the original symmetries and can be used
to define other BRST and conformally invariant actions with different values of
the anomaly coefficient, although the starting actions are related by canonical
transformations. At the classical level the new energy momentum tensors
differ from the   one of the the ghost action of 2-D supergravity  by   terms
which are   derivatives of $U(1)$ currents. As an interesting application, we
show    the relationship between the  ghost action
of 2-D supergravity and the action  of topological 2-D gravity in the gauge of
Labastida, Pernicci and Witten \cite{lpw}. Further field redefinitions permit
one to introduce the chiral actions  whose gauge variations reproduce the left
or right conformal anomaly with an arbitrary coefficient, that is  left-right
assymmetric anomaly compensating  actions, which can be combined to other
systems \cite{pol} \cite{yos} \cite{bbg}. All our formula are given for the
holomorphic sector and those of the  antiholomorphic sector are trivially
obtainable by conjugation. One has the amusing possibility of doing different
transmutations in the left and right sectors, starting from supergravity.

\section{Conformally invariant ghost system of 2-D supergravity}

The basic fields of 2-D supergravity are a zweibein and its supersymmetric
partner, the gravitino. It is possible to extract from these fields objects
which only depend on the superconformal classes, called moduli and supermoduli,
that is the Beltrami and superBeltrami parameters. In the holomorphic sector
one denotes them as $\m$ and $\a$. On a given Riemann surface, up to a
conformal
factor, the line element is simply $ds^2=|dz+\m d\z|^2$. General
super-reparametrizations transform  $\m$  and its supersymmetric partner $\a$
with the rules of a closed algebra, in relation with the superVirasoro algebra.
This algebra can thus be expressed as  a nilpotent BRST algebra
by changing the local diffeomorphism parameter into the anticommuting ghost
$\c$ and the local supersymmetry parameter  into the  commuting ghost $\g$.
The action of its graded differential BRST operator $s$ is defined as
\cite{bbgi}\cite{bbg}\be
\label{brssg} s\m&=& \pb\c+\c\pz\m-\m\pz\c-\demi \a\g \nn\\ s\c&=&  \c\pz\c-
\quart \g\g \nn\\ s\a&=& -\pb\g+\c\pz\a+\demi\a\pz\c+\m\pz\g-\demi\g\pz\m \nn\\
s\g&=&  \c\pz\g- \demi\g\pz\c \ee

There is a global $N=1$ supersymmetry inherent to the supergravity. Let $\s$
its generator, defined as  \be \s\m&=& \a \quad\quad \s\a =\pz \m \nn\\ \s\c&=&
\g \quad\quad \s\g =\pz c \ee

The way  the symmetries (anti)commute with the space derivatives is $\s d+d\s
=d
s+sd =0$, where $d=dz\pz+d\z\pb$. The consistency of the underlying symmetries
implies
 \be
\label{commut} s^2=0\quad\quad \s^2=\pz\quad\quad\s s+s\s =0  \ee

These formula can be unified in a superfield formalism. Let us interpret $\t$
as
the single Grassman variable of the $N=1$ supersymmetry in the holomorphic
sector and define the superfields \def\e{\epsilon ^z} \def\k{\kappa ^\t}
\def\demi{{1\over 2}} \def\quart{{1\over 4}} \def\C{C^z} \def\M{M^z}
 \be
\M&=&dz+\m d\z+\t\a d\z \nn\\ \C&=&\c+\t\g \ee
 One finds easily that   $\s$ is
the following graded differential operator
 \bq \s=\pt+\t\pz \eq
 and that the BRST equations
can   be written in superfield notations as
 \be
s\M&=&-d\C+\C\pz\M+\M\pz\C-\demi\s\C\s\M \nn\\ s\C &=& \C\pz\C-\quart{(\s\C
)}^2
\ee
One has a further simplifications  in the notation where the ghost
 number and the form degree are unified in a bigrading (to read the formula,
just expand them in form and ghost number).  Indeed the definition of the
operator $s$ can be written as   the more geometrical equation
 \def\e{\epsilon ^z} \def\k{\kappa ^\t} \def\demi{{1\over 2}}
\def\quart{{1\over 4}} \def\C{C^z} \def\M{M^z} \def\MM{\M+\C} \def\dd{d+s}
 \be
\label{brssggeo}
(\dd )(\MM) = (\MM)\pz(\MM) -\quart \s(\MM) \s(\MM) \ee
  Moreover one can also verify
\be
\label{brssggeoo}
 (\dd )\s(\MM)
= (\MM)\pz\s(\MM) -\demi \s(\MM)\pz(\MM)
 \ee
 which shows that $\s(\MM)$ has holomorphic weight $\demi$, as it should,
since it is the supersymmetric partner of the object $\MM$ which has weight
$1$.

 \section{The ghost action of 2-D supergravity in non trivial background
gauges}

Consider now the problem of gauge fixing a superconformal invariant action in
the gauge in which the supermetric of the    worldsheet is
conformally set equal to a given background supermetric, with
superBeltrami parametrization ${\m}_0$ and ${\a}_0$. In the holomorphic sector,
this generates the following Faddeev-Popov action
\be \I=\int d^2 z\ s(\b (\m
-{\m}_0)+\B (\a  -{\a}_0)) \ee
 $\b$ and $\B$ are antighosts with  the following BRST transformations
  \def\K{k_{zz}}
\def\G{k_{z\t}}
 \be s\b=\K\quad\quad s\K=0 \nn\\ s\b=\G\quad\quad s\G=0 \ee
Of course
the BRST operator $s$ does not act at this level on the background fields,
$s\mo=s\ao=0$.
The action is automatically BRST invariant, since $s^2=0$ on all
fields. Expanding the action from the definition of $s$ gives
\be \I=\int d^2 z &&\{\K
(\m -{\m}_0)+\G (\a  -{\a}_0)\nn\\ &&-\b(\pb\c+\c\pz\m-\m\pz\c-\demi \a\g)\nn\\
&&+\B(-\pb\g+\c\pz\a+\demi\a\pz\c+\m\pz\g-\demi\g\pz\m)\ \}\ee
Eliminating the Lagranger multipliers $\K$ and $\G$, one gets $\m=\mo$ and
$\a=\ao$ and thus
 \be
\I&=&
\int d^2 z\{ -\b(\pb\c+\c\pz\mo-\mo\pz\c-\demi \ao\g)\nn\\
&&\quad +\B(-\pb\g+\c\pz\ao+\demi\ao\pz\c+\mo\pz\g-\demi\g\pz\mo)\ \}
\ee
 In this equation, as a result of the gauge-fixing $\m$ and $\a$ have been set
equal to the classical backgrounds $\mo$ and $\ao$. Only the ghosts and
antighosts are dynamical  and the BRST symmetry operator $s$ has been changed
into $s_0$, which acts now in a non trivial way  on the background
metric and gravitino, with
\be
\I=
\int d^2 z\
(-\b s_0\mo+\B s_0\ao) \ee
where $s_0$ is defined as
 \be s_0\mo&=& \pb\c+\c\pz\mo-\mo\pz\c-\demi \ao\g \nn\\ s_0\c&=&  \c\pz\c-
\quart \g\g \nn\\ s_0\ao&=&
-\pb\g+\c\pz\ao+\demi\ao\pz\c+\mo\pz\g-\demi\g\pz\mo
\nn\\ s_0\g&=&  \c\pz\g- \demi\g\pz\c \ee
and
 \be s_0\b=s_0\B=0 \ee
 The new BRST invariance of the action, $s_0\I=0$,
  is   obvious from  $s_0^2=0$.

We   define  the $\s$ supersymmetry transformations of the antighosts as
 \be
\s\B=\b\quad\quad \s\b=\pz\B \ee
 which implies  that  \def\e{\epsilon ^z}
\def\k{\kappa ^\t} \def\demi{{1\over 2}} \def\quart{{1\over 4}} \def\C{C^z}
\def\M{M^z} \def\MM{\M+\C} \def\dd{d+s} \def\BB{B_{z\t}}
 \be \BB=\B+\t\b \ee
 is
a superfield  with $s_0\BB=0$ and
\be
\s\I=0
\ee
 The difference between the operators $s$ and
$s_0$ originates in the elimination of the Lagrange multiplier fields. From
now on, for the sake of notational simplicity, we will skip the index $"_0"$,
keeping in mind that the $\m$ and $\a$ are background fields defined on the
worldsheet.

The following expressions
 of the action $\I$ and the various anticommutation relations of $s_0$
and $\s$, similar to \ref{commut}, make obvious its invariances under $s$ and
$\s$ transformations  \be
\I&=&\int d^2 z\ s(\b  \m+\B  \a)= \int d^2 z\ s\s( \B  \m) =\int d^2 z d\t
s(\BB\M) \ee

\section{Conformal properties of  2-D supergravity BRST symmetry }

We now look for the transformations properties under superconformal
transformations of the field system discussed in the
previous section. \def\e{\epsilon ^z} \def\k{\kappa ^\t} \def\demi{{1\over 2}}
\def\quart{{1\over 4}} \def\C{C^z} \def\M{M^z} \def\MM{\M+\C} \def\dd{d+s}
\def\BB{B_{z\t}} \def\l{l^z} \def\ll{\lambda^\t}

We define "holomorphic superconformal transformations" by the action of the
following graded generator
 \def\e{\epsilon ^z} \def\k{\kappa ^\t}
\def\demi{{1\over 2}}
\def\quart{{1\over 4}} \def\C{C^z}
 \def\M{M^z} \def\MM{\M+\C}
\def\dd{d+s}
\def\BB{B_{z\t}}
\def\l{l^z} \def\ll{\lambda^\t}
 \def\d{\delta}
 \be
\label{defdelta}
\d\m&=&
\pb\l+\l\pz\m-\m\pz\l-\demi \a\ll \nn\\ \d\c&=&  \l\pz\c +\c\pz\l - \demi \ll\g
\nn\\ \d\a&=& -\pb\ll+\l\pz\a+\demi\a\pz\l+\m\pz\ll-\demi\ll\pz\m \nn\\
\d\g&=&  \l\pz\g-\demi\g\pz\l+c\pz\ll- \demi\ll\pz\c \ee
 To obtain the convenient grading, namely that
$\delta$ is odd,  we have "ghostified" the parameters of the transformations,
which means that the local diffeomorphism
 and supersymmetry parameter $\l$ and $\ll$ are respectively anticommuting and
 commuting.

When  $\l=0$ and   $\ll=constant$, the local
 symmetry defined from $\d$ reduces to the global symmetry with generator  $\s$
 that we  discussed in the previous section.
The $\d$ transformations form  a closed algebra. Moreover,
  by introducing the super-parameter
 \def\L{{\Lambda^{z}}}
\be \L=\l+\t\ll \ee
and using  the ghost form degree bigrading, we get for the $\d$
transformations,
 \be \d
(\MM)=-d\L+\L\pz(\MM)+ (\MM) \pz\L-\demi\s\L\s(\MM) \nn\\ \ee

The way $\d$ acts on the BRST transformed fields is instructive \footnote{To
simplify the formulae, one could   interpret $\L$ as a superghost. In this way
the differential operator $\d$ becomes a background BRST operator, with its own
ghost number, and one gets the unified equation  $(d+s+\d)(\MM+\L)=
 (\MM+\L)\pz(\MM+\L)-\quart\s(\MM+\L)^2 $, which makes particularly easy
the demonstration of most formulae.}. Using
 \be
 s(\MM)=-d(\MM)+ (\MM) \pz(\MM)
-\quart\s(\MM) \s(\MM) \nn\\ \ee  one gets after a simple computation
 \be \d
( s (\MM))= \L\pz s(\MM)+  s(\MM) \pz\L-\demi\s \L\s s(\MM) \ee This equation
shows that   the background superconformal graded operator $\d$ is compatible
with the BRST operator $s$, that is, $s$ and $\d$ anticommute,
\be \d s+s\d =0 \ee

In order that the   action $\I$ be invariant under the  transformations $\d$,
we
define the $\d$ transformation of the antighosts $\BB=\B+\t\b$   as
 \be
\d\BB=\L\pz\BB+\trois\BB\pz\L-\demi\s\BB\s\L \ee
that is in components
\be
\label{defdeltaa}
\d\b&=&\l\pz \b-2\b\pz \l+\trois\b\pz\ll+\demi\B\pz\ll \nn\\ \d\B
&=&\l\pz\B+\trois\B\pz \l-\demi \b\ll \ee
Indeed, this  definition implies
 \be \d
(s(\BB\M))=\pz(\ldots)+\s(\ldots) \ee
which proves
the $\d$ invariance of the supergravity ghost action
\be
\d\I=\d\int d^2zd\t s(\BB\M)=0
\ee

Another way to see the possibility of the $\d$ invariance of the action  is to
examine   $\I=\int d^2 z\ s(\b (\m -{\m}_0)+\B ( \a  -{\a}_0))$   before the
elimination of   the Lagrange multipliers  which are the BRST transformed of
the antighosts. Since $\m$ and ${\m}_0$ on the  one hand, $\a$ and $ {\a}_0$ on
the other hand, transform identically under $\d$, the form of  $\d(\m-{\m}_0)$
and $\d(\a-{\a}_0)$ is exactly what is needed to make possible $\d(\b (\m
-{\m}_0)+\B ( \a  -{\a}_0))=\pz(\ldots)$, which implies the $\d$ invariance of
the action. The  transformation laws under  $\d$ of the
Lagrange multipliers   must correspond to those of  the antighosts. Thus, all
relations of the type   \ref{commut} can be   extended in the antighost sector.
\def\L{{\Lambda^{z}}} \def\TT{{T_{zz}}} \def\GG{G_{z\t}}

The Ward identity of the $\d$ invariance permits a direct derivation of the
expression of   supersymmetry and energy momentum tensors $\GG$ and $\TT$,
defined as
 \be \TT={\d\I\over \d
\m} \quad\quad \GG={\d\I\over \d \a} \ee
The $\d$ invariance of the action means
\be
\label{slavnov} \int d^2 z(\TT\d\m+\GG\d \a + {\d\I\over \d \b}\d \b
+{\d\I\over
\d  \c}\d \c +{\d\I\over \d \B}\d \B +{\d\I\over \d  \g}\d \g) =0 \ee
 This equation
can be separated in  two identities, corresponding to independent   values of
the parameters $\l$ and $\ll$. Up to the equations of motion of propagating
fields, that is of the ghosts, \ref{slavnov} gives
\be \int d^2 z(\TT\d\m+\GG\d \a ) =0
\ee
Inserting the expression of $\d \m$ and $\d \a$ and using the fact that one
has identities true for all possible values of  the  parameters $\l$ and
$\ll$, we obtain
 \be
\label{wi} (\pb-\m\pz-2\pz\m)\TT- (\demi\a\pz+\trois\pz
\a)\GG&=&0\nn\\ (\pb-\m\pz-\trois\pz\m)\GG- \demi\a\TT &=&0 \ee
 These
Ward identities, valid in the presence of   general  values of the background
metric $\m$ and gravitino $\a$ of the worldsheet,  are the covariant
generalization of the   analyticity conditions of $\TT$ and $\GG$ in the
conformal gauge   $\m=\a=0$.

In superfield notations,   the "super energy-momentum tensor" is
 \def\L{{\Lambda^{z}}} \def\TT{{T_{zz}}} \def\GG{G_{z\t}}
\def\KK{\tilde{T}_{z\t}}
\be \KK=\GG+\t\TT \ee
 One easily verifies
\be \KK= {\d\I\over \d  \M}\int d^2 z d\t (\BB s\M)
= \C\pz\BB+\trois\BB\pz\C-\demi\s\C\s\BB \ee
and another way to write the Ward identity \ref {wi}  is
  \be
\label{swi}(\pb-\M\pz-\trois\pz\M-\demi\s\M\s)\KK=0 \ee

 Since the action is linear in $\m$ and $\a$, $\TT$ and $\GG$
are
independent on these fields. Let us check that our definitions truly give
the known definitions of $\TT$ and
$\GG$ in the conformal gauge for which $\m=\a=0$.
One has indeed
\be
\label{emt}
 \TT&=&{\d\I\over \d  \m}=\int d^2z (\b{\d \over \d  \m}s\m+\B {\d
\over \d \m }s\a )
\nn\\ &=& \c\pz\b-2\b\pz\c+\demi\g\pz\B+\trois\B\pz\g \ee
 and
\be
\GG={\d\I\over \d  \a}={\GG}^++{\GG}^- \ee
where
\be {\GG}^+&=&\int d^2z  \b{\d  \over \d
\a}s\m =\demi\b\g \nn\\ {\GG}^-&=&\int d^2z  \B{\d \over \d  \a}s\a =
\demi\B\pz\c+\pz(\B\c) \ee

 If one switches to a classical Hamiltonian formalism, the form of the action
indicates that   the antighost  field
  is the conjugate momenta of the ghost field operator.
Moreover the action of  a  conformal
super-reparametrization $\d$ with super-parameter $\L$
on any given dynamical field  $X$ can be written as
$\d X={1\over{2\pi i}}\{  \oint
dz d\t\L  \KK  ,  X  \}_\pm$, where the  anti-bracket $\{\ ,\ \}_+$
occurs if $X$ has an odd grading. The group structure
 of the  transformations $\d$ implies
the anticommutation relation
 \be \{\oint dz d\t\L  \KK, \oint dz  d\t{\L  }'\KK\}_+=\nn\\ \oint dzd\t(\L
\pz{\L  }' +\trois \L  \pz{\L  }'-\demi \s\L  \s{\L  }' )\KK)  \ee
where
$\L=\l+\t\L$  and
 ${\L}'={\l}'+\t{\L}'$ stand for the super-parameters of  two $\d$
transformations.

One may decompose the   last equation by projection over
 the various  possibilities of the component content of
   the super-parameters $\L$. This gives the graded commutation relations of
the
holomorphic N=1 superconformal algebra
\be \{ \TT ,\TT \}   \sim    {\TT } \quad \{ {\GG}
,{\TT}\}     \sim   {\GG }
 \quad
\label{sst}\{ {\GG} ,{\GG}\} _{+}   \sim   {\TT }
 \ee
 where we do not make explicit the structure coefficients for the sake of
notational simplicity. Since   $\{ {\GG}^+ ,{\GG}^+\} _{+} =$$\{ {\GG}^-
,{\GG}^-\} _{+}=0$, one has   \be \{ {\GG}^- ,{\GG}^+\} _{+}   \sim   {\TT }
 \ee

For any given value of the background fields $\m$ and $\a$,
one has an obviously classically conserved abelian current, the ghost current
\be
 J_z^{ghost}=\b\c+\B\g \ee
However, for $\a=0$  one can observe that the two currents $\b\c$ and $\B\g$
are separately conserved due to the vanishing of   mixing terms in the
invariant action, $\b\a\g$ and $\B(\c\pz\a+\demi\a\pz\c)$. Moreover,
the nilpotent
transformation $\s^+$ associated to ${\GG}^+$, namely
 \be \s^+ \c=\g\quad\quad
\s^+\g=0\nn\\
 \s^+\B=\b\quad\quad \s^+\b=0 \ee
  that is $\s^+=\pt$ in superfield notation, is a symmetry of the action if and
only if
  \be
\label{condi}
\a=0\quad\quad
 \pz \m=0 \ee
But in this case     the action is also invariant under the action of
${\GG}^-$ since it is generally invariant under the  fundamental symmetry
associated  to  ${\GG}={\GG}^++{\GG}^-$. Thus, when the background gauge is
restricted as   in \ref{condi},
 and in particular in the
conformal gauge case $\m=0$,   one has two supersymmetries of the
action associated to the generators ${\GG}^+$ and ${\GG}^-$. This gives
  the known  $N=2$ superconformal supersymmetry, for which   $J_z^+=\b\c  $
or $J_z^-= \B\g $  plays the role of the abelian part, while
${\GG}^+$ and ${\GG}^-$ are the two fermionic generators \cite{fms}. From our
point
 of view this
symmetry appears as rather accidental, the basic symmetry being the
one associated to  ${\GG} $, $\TT$ and  $ J_z^{ghost}$.

 After quantization of  the
 action, which means either doing path integral over the  dynamical fields or
changing the fields into operators and Poisson bracket into commutators, the
anomaly can be understood as a consistent term generated by loop
corrections which can be substituted to zero at the right hand side of the
superconformal Ward identity \ref {wi} or \ref {swi}. After a quick look to the
structure equations one finds that the consistency of the symmetry equations
implies that the Ward identity
 \ref {wi} or \ref {swi} can be made anomalous only under
the following form
 \be
\label{swib}(\pb-\M\pz-\trois\pz\M-\demi\s\M)\KK&=&{{ \it \bf
c}}\pz(\pt+\t\pz)\pz\M\nn\\ &=&{ \it \bf c}(\pz^2\a+\t\pz^3\m)
\ee
 The value of the
coefficient ${{ \it \bf c}}$ of the anomaly
depends on the explicit form of $\TT$. It must  be computed at the quantum
level, using one of the many methods available. One gets ${{ \it \bf
c}}=-26+11=-15$ for the case of the 2-D supergravity ghost action.

  \section{Transmutations to other conformal theories}

The ghost action which stems from the
conformal gauge fixing to a non trivial background structure  of 2-D
supergravity possesses the    super-reparametrization invariance
\ref{defdelta} \ref{defdeltaa} in addition to  its BRST invariance. It is thus
conceivable to think of other conformal theories that one would obtain by
introducing new fields which are composites of the original ghosts of the
supergravity, obtained by products or more subtle combinations of these fields,
with the possibility of a dependence on the backgrounds fields $\m$ and $\a$,
so
that one has a priori many options to choose from for the holomorphic weights
as  well as the statistics. The new fields would be well defined under the
local background super-reparametrization invariance since $\d$ acts as
differential operator. The theories stemming from BRST invariant actions built
from these fields would also have conformal properties, since $s$ and $\d$
anticommute. The values of their anomaly coefficients would differ, since the
weights of the fields would be different, but their Lagrangians would be
related by canonical changes of field variables, as well as their  classical
energy momentum tensors, BRST Noether currents and all other conserved
currents.

To understand this construction, we will explain how the action of the
topological 2-D gravity in the type of gauge used by Labastida Pernicci and
Witten \cite{lpw} is simply related, through a canonical change of variable, to
the one of 2-D supergravity discussed in the previous section.
 Let us redefine
\def\PP{{\Psi}^z_\z} \def\PH{\Phi^z } \def\PHb{{\bar\Phi}_{zz} }
\def\ne{\bar\eta_{zz} } \def\g{\gamma^\t}
\be
\label{cv} \PP&=&-\demi \g \a \nn\\
 \PH &=&-\quart \g\g\nn\\
\PHb &=&-2 \B{\g}^{-1}   \ee
while we keep unchanged $b_{zz}$ and $\c$. It is easy to check that this
 change of variable is canonical, which means that the Poisson bracket of the
redefined fields $\PHb$ and $\PH$ is equal to that of the conjugate fields $\B$
and $\g$.

To determine the BRST algebra of the redefined fields, we insert in   the basic
supergravity BRST equations
\ref {brssggeo} and \ref{brssggeoo} the   change of variable \ref{cv},
observing that it means $\PP+\PH=-\quart\s(\MM)\s(\MM)\mid_{\t=0}$.
One obtains
\def\MMM{{dz+\m d\z+\c}} \be (d+s)(\MMM)&=&(\MMM)\pz(\MMM)+\PP+\PH
\nn\\ (d+s)(\PP+\PH)&=&(\MMM)\pz(\PP+\PH)+(\PP+\PH)\pz(\MMM) \nn\\ \ee
Expanded in ghost number, these equations mean
\be s\m&=&\PP+\pb\c+\c\pz\m-\m\pz\c
\nn\\ s\c &=& \PH+\c\pz\c \nn\\
s\PP&=&-\pb\PH+\PH\pz\m-\m\pz\PH+\c\pz\PP+\PP\pz\c \nn\\
 s\PH&=&  \c\pz\PH-\PH\pz\c \ee
 One recognizes the    BRST operator of topological 2-D gravity,
as expressed in \cite{bsgr}. Thus the BRST symmetry of 2-D supergravity has
been transformed into that of topological 2-D gravity by the   change of
variables \ref{cv}. It is  quite interesting that the  combination of the
anticommuting physical gravitino $\a$ with its commuting ghost $\g$
 produces the anticommuting   ghost $\PP$, that is
the topological ghost partner of the Beltrami variable $\m$, while the square
of
$\g$ produces the topological ghost of ghost $\PH$.

Consider now the BRST invariant action
 \def\II{\I_{\rm top}}
\be \II=\int d^2z s\{ \b(\m-{\m}_0)+\PHb(\PP+\demi\ao\g)
\} \ee
 The
action of $s$ on the antighosts is
 \be s\b=\lambda_{zz} \quad\quad
s\lambda_{zz}=0\nn\\ s\PHb=\ne \quad\quad s\ne=0 \ee
 Expanding the action $\II$
we see  that $\lambda_{zz}$ plays the role of  a commuting Lagrange multiplier
for the condition
 $\m=\mo$ while $\b$ is an
 anticommuting Lagrange multiplier for the condition
$\PP=-\pb\c-\c\pz\mo+\mo\pz\c+\demi\ao\g$. Thus, only the second term of $\II$
gives rise to a dynamical part. If furthermore we choose the background
gravitino $\ao=0$, what remains  of the action after  the elimination of the
Lagrange multiplier  fields is
 \be \label{acttopg}\II=\int d^2z\{   &&\ne (\pb-\mo\pz+\pz\mo)\c \nn\\ &&-
\PHb
(\pb-\mo\pz+\pz\mo)(\PH+\c\pz\c)\ \}  \ee

It is easy to verify that $\II$ is invariant under the  background symmetry
 $\d$. The expression of  $\d$ can be obtained from that we derived in 2-D
supergravity, eqs. \ref{defdelta} \ref{defdeltaa}, through our changes of
variables.
 The transformations laws of the redefined fields under  $\d$
 permits of course to
verify  that their  conformal weights  are truly what is
indicated by our indices.

As far as its interpretation is concerned, the  action $\II$ is of the
Labastida-Pernici-Witten type \cite{lpw}, expressed in the Beltrami
parametrization. It can be considered as a holomorphic gauge fixed version of
the topological invariant $\int d^2x {\sqrt g}R$, by mean of the gauge
condition
$\m=\mo$ \cite{bsgr}.
 Before
quantization, the energy momentum tensor, can be equivalently computed   by
differentiation of the topological gravity action
\ref{acttopg}  with respect to   $\mo$,  or by doing the  changes of field
variables  \ref{cv} in the energy momentum tensor of 2-D
supergravity \ref{emt}. If the fields are
identified, the two energy momentum tensors differ  by a term of the type
$\pz(a\b\g+a'\B\c)$. The BRST symmetry Noether currents   of both
theories are   also related by the canonical changes of variables. One still
has
the Ward identities  \ref {wi} with the same possible anomalous term as in \ref
{swib}. The value of the anomaly coefficient
${{ \it \bf c}}$, not predicted from classical
considerations,  is now zero, due to an exact compensation between the
contributions of commuting and anticommuting ghosts.   The discussion about the
$N=2$ conformal supersymmetry of the redefined action can be repeated exactly
as
in the last section, when the background is such that $\mo =0$.

The mechanism presented here, which allows the
change of the holomorphic weight and statistics of the propagating fields, by
mean of canonical field redefinition  involving the gravitino and its ghost,
that is the spin structure of the wordsheet,  indicates a possibly deep
relationship between two   actions with    different physical interpretations.
The canonical changes on field variables,
$ (\ne,\c)=(\b,\c)$ and $ (\PHb,\PH)=(-2\B{\g}^{-1},-\quart \g\g)$ connects
the symmetries, the field equations, the energy momentum tensor  and other
conserved currents of both actions. On the other hand, the anomaly coefficient,
whose computation relies on quantum effects    changes. It is   intriguing to
find out whether the mechanism which governs the change of the values of the
anomaly  coefficient is linked to an ordering problem or to the singularity
occurring at $\g=0$ in the field redefinitions.

Having seen the existence of the transmutation mechanism on a given example, we
may go
  further and     consider the possibility of introducing other $b-c$ systems,
the root being   the ghost system of 2-D supergravity that one transforms by
canonical field redefinitions.  We   introduce at the classical level a
pair of new fields by the following canonical change of variables
 \be
\label{cvli} \exp -{L\over a}&=&\PH+\c\pz\c\nn\\ M&=&-{ \PHb\over a}\exp -
{L\over a} \ee
 $a$ is an arbitrarly fixed real number. In terms of the redefined
fields, the action is  \be
\label{la}
\II=\int
d^2z\{  \ne (\pb-\mo\pz+\pz\mo)\c  - M  ( \pb L-\mo\pz  L-a\pz\mo)\}
  \ee
 The BRST symmetry of the action has been changed into
\be
\label{brsla}
 s\mo&=& \pb\c-\mo\pz\c+\c\pz\mo
 \nn\\
  s\c&=&\c\pz\c
\nn\\
 sL&=&c\pz L-a\pz \c
\nn\\
 s\ne&=&0\nn\\
 sM&=&0
\ee
 From a mathematical point of view, the meaning of the fields $L$ and $M$ is
not
clear when $a$ is not integer or half integer. However, we have the
background
conformal symmetry
 \be \delta L&=&\l \pz L-a \pz \l \nn\\ \delta M&=& \pz (\l M) \ee
 which leaves
invariant the action. We may think to define the nature of the fields from
these equations. Moreover, the physical meaning of the  action \ref{la} is
quite
clear. The ghost part  $\ne (\pb-\mo\pz+\pz\mo)\c$  means that the Beltrami
parameter of the worldsheet has been set equal to the background value  $\mo$
in
a BRST invariant way. Interpreting the field $M$ as a Lagrange multiplier, the
part $M  ((\pb-\mo\pz )L-a\pz\mo)$ means that one has a field $L$ which
satisfies the constraint \cite{pol}\cite{yos}\cite{bbg}
 \be
  (\pb-\mo\pz )L=a\pz\mo  \ee
We have the following property
 \be s(\pz
L(\pb-\mo\pz )L-2\pz \mo\pz L)=a \pz\c {\pz}^2\mo \ee
which holds true  as a consequence of \ref{brsla}, independently, of the
constraint $ (\pb-\mo\pz )L=a\pz\mo $. If, moreover, this constraints is
satisfied, we have
  \be s(\pz \mo\pz L)=-a\pz\c {\pz}^2\mo \ee
  Thus, the constrained field $L$   can be used to compensate the conformal
anomaly, since the gauge variation of the action $\pz \mo\pz L$ is
proportional to the conformal anomaly $\pz\c {\pz}^2\mo$. Moreover, at the
quantum level, we have an admissible invariant counterterm, of the
cosmological term type, which can  be added to the action, \be
 \I_{ct}=cte\int d^2z \ \exp -{L\over a}
\ee
since the  $s$ and $\delta$ variation of the integrand  is a pure derivative.
All these properties indicate that we may call the      field  $L$ "half
a Liouville field"  as in \cite{bbg}.

 To summarize, through our successive changes of field variables, we have
reached   the action
 \be
\label{brswz}
 \I_{WZ}= \int d^2z
 \{\  &&\ne (\pb-\mo\pz+\pz\mo)\c)\nn\\ &&- M  ((\pb-\mo\pz )L-a\pz\mo)+b\pz
\mo\pz L\}\ee
where $b$ is an arbitrarly given real number, independent of a. The energy
momentum tensor is \be
T_{zz}=
{
{\delta \I_{WZ}}
\over
{\delta \mo}
}
=
-\ne \pz\c-\pz(\ne \pz\c)
+ M   \pz  L-a\pz  M+b\pz^2 L
\ee
An easy computation shows that
the  contribution of
all the fields to the anomaly coefficient is
 \be {\it \bf c} =-26+12ab+1 \ee
If one is interested to get a model with a vanishing
 anomaly coefficient,
 ${\it \bf c}=0$,  there are many choices for the values of the pairs $a,b$.
All
possible values of $a$ are admissible, provided
\be
b={25\over{12a}}
\ee

As far as observables are concerned, one relies on the criteria of
BRST invariance for their selection \cite{ver}. But since the BRST symmetry is
the
same in all these models, up to field redefinitions, we expect a universality
in
the definition of the products of fields of which one should take the
expectation
values. As an example, in the topological phase,
we have the BRST-exact cocycle $\Phi^z+\c\pz\c$ with ghost number two as an
observable, which corresponds in the  Liouville phase   to the BRST invariant
"cosmological  term"  $exp -{L\over a}$, through the redefinition \ref{cvli}.
Notice that in the topological phase a supersymmetry breaking mechanism should
occur in order  that $<\Phi^z+\c\pz\c>$ be not zero, due to    its
ghost number.  The values of the expectation values of these physical
operators,
as well as the methods  of computation, should  differ when one goes from one
model to the other.  It
is   quite interesting that the reduced cohomology introduced at the algebraic
level in \cite{bow},
 can be directly deduced from the BRST charge corresponding to the
action \ref{brswz}.
\\
{\bf {Acknowledgments}} I thank J.L. Gervais, M. Picco and P. Windey for many
instructive discussions related to  the subject of this work.



 \end{document}